\documentclass[11pt,reqno]{article}

\usepackage{amsmath}
\usepackage{amssymb}

\newcommand{\metric}{\mathrm{g}}

\begin{document}

\title{Fluctuation Spectrum from a Scalar-Tensor Bimetric Gravity Theory}

\author{
\vspace{0.2cm}
M.~A.~Clayton\thanks{michael.clayton@acadiau.ca} \\
\vspace{0.2cm}
\textit{Department of Physics, Acadia University, Wolfville, Nova Scotia, B0P 1X0, Canada}\\
\vspace{0.2cm}
and\\
\vspace{0.2cm}
J. W.~Moffat\thanks{john.moffat@utoronto.ca}\\
\textit{Department of Physics, University of Toronto, Toronto,
Ontario M5S 1A7, Canada, and} \\
\textit{Perimeter Institute for Theoretical Physics, Waterloo, Ontario N2J
2W9, Canada}
}

\date{\today\\
\vskip 0.5cm
PACS: 98.80.C; 04.20.G; 04.40.-b \\
Cosmology, causality, inflation, alternative theories of gravity
}

\maketitle

\begin{abstract}
Predictions of the CMB spectrum from a
bimetric gravity theory (BGT)~\cite{Clayton+Moffat:2001} are presented. The initial
inflationary period in BGT is driven by a vanishingly small speed of gravitational
waves $v_g$ in the very early universe. This initial inflationary period is
insensitive to the choice of scalar field potential and initial values of the scalar
field. After this initial period of inflation, $v_g$ will increase rapidly and the
effects of a potential will become important. We show that a quadratic potential
introduced into BGT yields an approximately flat spectrum with inflation parameters:
$n_s=0.98$, $n_t=-0.027$, $\alpha_s=-3.2\times 10^{-4}$ and
$\alpha_t=-5.0\times 10^{-4}$, with $r\ge 0.014$.
\end{abstract}

\section{Introduction}

Recently we have introduced a cosmological
model~\cite{Clayton+Moffat:1999a,Clayton+Moffat:2001}, based on a
scalar-tensor bimetric gravitational theory (BGT) in which the
relative propagation velocities of gravitational and matter
disturbances is determined dynamically. In the resulting
diffeomorphism invariant theory there is a prior-geometric
relationship between two metrics that involves the gradient of a
scalar field. Similar models based on a vector
field have been published~\cite{Clayton+Moffat:1999,Clayton+Moffat:2000}.
Due to its role in producing a theory with multiple light cones (or
birefringent spacetimes), the scalar field is referred to as the
``biscalar'' field.  Depending on the choice of frame, spacetimes
may either be viewed as having a fixed speed of light and a
dynamically-determined speed of gravitational disturbances, or as
having a fixed speed of gravitational waves and a dynamical speed
of
light~\cite{Clayton+Moffat:1999a,Clayton+Moffat:2001,Moffat:2002}.

The BGT cosmological model was shown to possess solutions that had
sufficient inflation to solve the horizon and flatness problems,
and in order to bring it in line with current observations on the
CMB radiation, we now demonstrate that it is capable of generating
an approximately scale invariant primordial scalar perturbation spectrum.
To accomplish this we will adopt the standard scenario from slow-roll
inflation models, assuming an adiabatic quantum vacuum for the
perturbative scalar and tensor modes.  To accomplish this and at
the same time retain the usual picture of (pre-)heating following
inflation~\cite{Brandenberger:2000}, we will introduce a mass for
the biscalar field.

While this would seem to be a small deviation from slow-roll
inflation, because the speed of gravitational wave propagation is
less than the speed of light during inflation, the physics of the
tensor modes is altered. In particular, there is {\it no simple
consistency relation between the scalar mode spectral index and
the ratio of tensor and scalar spectra}. At the same time, the
biscalar field potential is strongly suppressed during an earlier
period of inflation, much as if a coupling constant ``ran'' to
zero in the very early universe. So, unlike standard slow-roll
inflation scenarios, it would be difficult to argue that any
particular value for the biscalar field (or the biscalar field
potential energy) in the very early universe is ``unnatural''.

Following this initial period of inflation, the biscalar field
potential will become important, and as the biscalar field
approaches the bottom of the potential, the universe is capable of
going through another period of inflation, resembling that of
chaotic inflation models~\cite{Linde+etal:1994,Linde:1994}. It is
during this period that we expect that the primordial seeds of
structure formation are generated from the quantum vacuum of the
biscalar field $\phi$, the results of which we describe in this
paper.

To begin with, we will briefly review the structure of the BGT
model in Section~\ref{sect:model intro}. Then, we shall examine
the very early universe in Section~\ref{sect:cosmology}, and
determine the conditions under which the model possesses inflating
solutions with a constant speed gravitational wave propagation
$v_g<c$.  In Section~\ref{sect:parameters}, we determine the
inflation parameters for a model in which vacuum fluctuations of
the biscalar field $\phi$ produce the primordial seeds of the CMB
spectrum, and constrain the model parameters to give the observed
density profile $\delta\rho/\rho\sim 10^{-5}$. We end in
Section~\ref{sect:conclusion}, with concluding remarks.

\section{The model}
\label{sect:model intro}

The model that we introduced
in~\cite{Clayton+Moffat:1999a,Clayton+Moffat:2001} consisted in a
minimally-coupled, self-gravitating biscalar field $\phi$ coupled to matter
through what we refer to as the ``matter metric'':
\begin{equation}\label{eq:bimetric}
\hat{\metric}_{\mu\nu}=\metric_{\mu\nu}+B\partial_\mu\phi\partial_\nu\phi.
\end{equation}

The action takes the form
\begin{equation}
\label{action}
S=S_{\rm grav}+S_{\phi}+\hat{S}_{\rm M},
\end{equation}
where
\begin{equation}
S_{\rm grav}=-\frac{1}{\kappa}\int d\mu (R[g]+2\Lambda),
\end{equation}
$\kappa=16\pi G/c^4$, $\Lambda$ is the cosmological constant, and
we employ a metric with signature $(+,-,-,-)$.  We will write, for
example, $d\mu=d^4x\,\sqrt{-g}$ and $\mu=\sqrt{-g}$ for the metric
density related to the gravitational metric $g_{\mu\nu}$, and
similar definitions of $d\hat{\mu}$ and $\hat{\mu}$ in terms of
the matter metric $\hat{g}_{\mu\nu}$. A useful result for two
metrics related by~\eqref{eq:bimetric} is that the determinants
are related by:
\begin{equation}
\mu = \sqrt{K}\hat{\mu},
\end{equation}
where
\begin{equation}\label{eq:K definition}
 K=1-B\hat{\metric}^{\mu\nu}\partial_\mu\phi\partial_\nu\phi.
\end{equation}

The minimally-coupled biscalar field action is given by
\begin{equation}
S_{\rm \phi}=\frac{1}{\kappa}\int d\mu\,
\Bigl[\frac{1}{2}g^{\mu\nu}\partial_\mu\phi\partial_\nu\phi-V(\phi)\Bigr],
\end{equation}
where the scalar field $\phi$ has been chosen to be dimensionless.
The gravitational field equations can be written as
\begin{equation}\label{eq:Einsteins eqns}
 G^{\mu\nu}=\Lambda\metric^{\mu\nu}
 +\frac{\kappa}{2}\biggl(T^{\mu\nu}_\phi
 +\frac{1}{\sqrt{K}}\hat{T}^{\mu\nu}\biggr),
\end{equation}
where $\hat{T}^{\mu\nu}$ is the matter energy-momentum tensor and
\begin{equation}
T^{\mu\nu}_\phi=\frac{1}{\kappa}\Bigl[
\metric^{\mu\alpha}\metric^{\nu\beta}\partial_\alpha\phi\partial_\beta\phi
-\tfrac{1}{2}\metric^{\mu\nu}\metric^{\alpha\beta}\partial_\alpha\phi\partial_\beta\phi
+\metric^{\mu\nu}V(\phi) \Bigr].
\end{equation}
The field equations for the scalar field, written here in terms of matter
covariant derivatives $\hat{\nabla}_\mu$, are given by
\begin{equation}\label{eq:scalar FEQ}
\bar{\metric}^{\mu\nu}\hat{\nabla}_\mu\hat{\nabla}_\nu\phi+KV^\prime
[\phi]=0.
\end{equation}
In the latter, we have defined the biscalar field metric $\bar{g}^{\mu\nu}$ by
\begin{equation}\label{eq:scalar metric}
 \bar{\metric}^{\mu\nu} = \hat{\metric}^{\mu\nu}
 +\frac{B}{K}\hat{\nabla}^\mu\phi\hat{\nabla}^\nu\phi
 -\kappa B\sqrt{K}\hat{T}^{\mu\nu}.
\end{equation}

The matter model is constructed solely using the
metric~\eqref{eq:bimetric}, resulting in the identification of
$\hat{\metric}_{\mu\nu}$ as the metric that provides the
arena on which matter fields interact. Therefore the
conservation of energy-momentum takes on the form:
\begin{equation}\label{eq:matterconservation}
\hat{\nabla}_\nu\Bigl[\hat{\mu}\hat{T}^{\mu\nu}\Bigr]=0,
\end{equation}
as a consequence of the matter field equations
only~\cite{Clayton+Moffat:1999a}.  It is the matter covariant
derivative $\hat{\nabla}_\mu$ that appears here, which is the
metric compatible covariant derivative determined by the matter
metric: $\hat{\nabla}_\alpha\hat{\metric}_{\mu\nu}=0$.
In this work, we will assume a perfect fluid form for the matter
fields
\begin{equation}
 \hat{T}^{\alpha\beta}=
 \Bigl(\rho+\frac{p}{c^2}\Bigr)\hat{u}^\alpha\hat{u}^\beta
 -p\hat{\metric}^{\alpha\beta},
\end{equation}
with $\hat{\metric}_{\mu\nu}\hat{u}^\mu\hat{u}^\nu=c^2$.

\section{Cosmology}
\label{sect:cosmology}

We will now specialize to a cosmological setting, imposing
homogeneity and isotropy on spacetime and writing the matter
metric in comoving form as
\begin{equation}
 \hat{\metric}_{\mu\nu}=\mathrm{diag}(c^2,-R^2(t)\gamma_{ij}),
\end{equation}
with coordinates $(t, x^i)$ and spatially flat ($k=0$) $3$-metric
$\gamma_{ij}$ on the spatial slices of constant time and $c$ is
the constant speed of light. In the following, we shall restrict
our attention to the variable speed of gravitational waves
frame~\cite{Moffat:2002}. The matter stress-energy tensor (using
$\hat{u}^0=c$) is given by
\begin{equation}
 \hat{T}^{00}=\rho,\quad
 \hat{T}^{ij}=\frac{p}{R^2}\gamma^{ij}.
\end{equation}
This leads to the conservation laws (an overdot indicates a
derivative with respect to the time variable $t$, and
$H=\dot{R}/R$ is the Hubble function):
\begin{equation}\label{eq:cosm cons}
\dot{\rho}+3H\Bigl(\rho+\frac{p}{c^2}\Bigr)=0.
\end{equation}

It is useful at this point to introduce the following quantities
derived from the constant $B$ which will appear throughout this
work:
\begin{equation}\label{eq:definitions}
H_B^2=\frac{c^2}{12B},\quad
\rho_B=\frac{1}{2\kappa c^2 B}.
\end{equation}
The latter is obtained by requiring that $H_B^2=\frac{1}{6}\kappa
c^4\rho_B$, and we have defined the dimensionless parameter:
\begin{equation}
 \xi_B^2 =\frac{\ell_P^2}{B},
\end{equation}
where the Planck length is defined by $\ell_P=\sqrt{\hbar G/c^3}$.
In~\cite{Clayton+Moffat:2001}, we described the parameters
required to generate sufficient inflation from a finite time after
the ``quantum gravity epoch''.  Just after this, spacetime
contained an approximately constant energy density $\rho_B$
defined by~\eqref{eq:definitions}. This motivated the
identification $\rho_B\approx \rho_P=c^5/(\hbar G^2)$, which would
correspond to $\xi_B^2=32\pi$.

From~\eqref{eq:K definition} we have
\begin{equation}\label{eq:K cosm}
 K=1-\frac{\dot{\phi}^2}{12H_B^2},
\end{equation}
and using the relation~\eqref{eq:bimetric}, the gravitational
metric is found to be
\begin{equation}\label{eq:grav metric}
 {\metric}_{\mu\nu}=\mathrm{diag}(v^2_g,-R^2\gamma_{ij}),
\end{equation}
where $v_g=cK^{1/2}\le c$ is the gravitational wave speed.  From the
definition~\eqref{eq:scalar metric} we find
\begin{equation}\label{eq:g bar}
 \bar{\metric}^{\mu\nu}
 =\mathrm{diag}\Bigl[\frac{1}{v_g^2}\Bigl(1-K^{3/2}\frac{\rho}{2\rho_B}\Bigr),
 -\frac{1}{R^2}\gamma^{ij}\Bigl(1 + \sqrt{K}\frac{\rho}{6\rho_B}\Bigr)\Bigr].
\end{equation}

The Friedmann equation ($R_{00}+Kc^2\gamma^{ij}R_{ij}$) is given
by
\begin{equation}\label{eq:Fried eqn}
 H^2 +\frac{kc^2K}{R^2} =
 \frac{1}{3}Kc^2\Lambda
 +H_B^2(1-K)
 +2H_B^2 K V_B
 +\frac{1}{6}\kappa c^4 K^{3/2}\rho.
\end{equation}
In~\cite{Clayton+Moffat:1999a,Clayton+Moffat:2001} we showed that very soon after an initial period of inflation during which $\sqrt{K}\rho\approx 2\rho_B$, we will have $\rho\ll \rho_B$ and so we will have $\bar{\metric}_{\mu\nu}\approx \metric_{\mu\nu}$.
In this limit, the biscalar field equation~\eqref{eq:scalar FEQ}
can be written in terms of $K$ as
\begin{equation}\label{eq:K dot}
 \dot{K}
 -6K(1-K)H
 -2 K^2 \dot{V}_B
 =0,
\end{equation}
where we have introduced the dimensionless potential: $V_B=BV$ and $\dot{V}_B=V^\prime_B\dot{\phi}$.
From here it is clear that while $K\approx 0$ the effect of the biscalar field potential is suppressed, during which time we have a second period of inflation as described previously~\cite{Clayton+Moffat:2001}.

If we want to understand the quantum origins of the CMB spectrum
and at the same time realize (pre-)heating as the decay of
oscillations of the biscalar field about a minimum of its
potential, then we must add a nontrivial potential to the model.
Rolling towards the minimum of this potential will cause inflation
similar to that appearing in chaotic inflation scenarios, during
which time modes of interest are being generated. The biscalar
field $\phi$ therefore provides both a mechanism to produce
inflation as the speed of gravitational waves $v_g\rightarrow 0$,
and in addition the fluctuations of $\phi$ on this inflating
background produces the CMB spectrum.This is the scenario
developed in this work.

An alternative scheme would be to separate inflation from the
production of the CMB spectrum. Assuming that inflation is
produced as described in~\cite{Clayton+Moffat:2001}, we could
introduce a second scalar field $\psi$ in the matter model, with
quantum fluctuations of $\psi$ generating the primordial quantum
fluctuations that determine the CMB spectrum after decoupling.
Since $v_g\rightarrow 0$ during inflation, quantum effects of the
biscalar and gravitational field would be suppressed, allowing
these scalar matter modes to dominate. This mechanism would be
independent of a biscalar field potential and does not depend on
the initial values of $\phi$, and so we anticipate that the
difference between the initial BGT inflation and the standard
slow-roll inflation mechanism can remove the severe fine tuning
problem of the latter models, caused by an unnaturally small
coupling constant $g\sim 10^{-13}-10^{-14}$ required to fit the
observed density profile, $\delta\rho/\rho\sim 10^{-5}$. We leave
the details of such a model to a later publication.

\subsection{The Very Early Universe}
\label{sect:very early}

We have previously demonstrated that if the potential $V(\phi)$ is
not important in the very early universe, then the field equations
require that an infinite amount of inflation takes place as
$R\rightarrow 0$~\cite{Clayton+Moffat:2001}. In these solutions we
found that $R\sim \exp(H_B t)$ (which is the origin of the
definitions~\eqref{eq:definitions}), and $K\sim R^6$ or $K\sim
R^8$.  In either case, since $K\rightarrow 0$ implies that
$\phi\rightarrow \mathrm{const.}$ as $R\rightarrow 0$, the fact
that the biscalar field potential enters into~\eqref{eq:Fried eqn}
and~\eqref{eq:K dot} multiplied by a power of $K$ implies that its
contribution to both will be vanishingly small.  Clearly this does
not rule out other possibilities (for example, that $V\gtrsim
1/R^{12}$ as $R\rightarrow 0$), but it is a good indication that
such behavior is likely in the very early universe. The result
that \textit{the onset of the initial inflationary period is
essentially independent of the potential} $V_B$, demonstrates an
important difference of BGT from the standard inflationary
models~\cite{Brandenberger:2000}, based on a slow-roll
approximation for the inflaton potential.

At the same time, the matter contribution to the Friedman equation
is scaled by a factor of $K^{\frac{3}{2}}$ with
$K^{\frac{3}{2}}\rho\approx \mathrm{constant}$ during this period.
In this way the effect of matter ``saturates'', and increasingly
large mass densities do not cause increasingly large curvatures;
an intriguing result. Given this, it is likely that this model
will have something to say about primordial black hole production.
Another intriguing aspect of the vanishingly small speed of
gravitational waves $v_g$ is that the exchange of gravitons
between two gravitationally interacting bodies is suppressed in
the very early universe, so that gravitating particles become
effectively asymptotically free. This could have important
consequences for quantum gravity.

We can imagine a scenario in which a more fundamental quantum
theory singles out an initial state with $K\approx 0$, so that the
biscalar field $\phi$ is initially rolling at a fixed rate
$\dot{\phi}=\sqrt{12}H_B$ but of no particular value, for
effectively there is no potential energy associated with it.
Different regions of the universe undergo the same type of early
inflation, regardless of the value of the biscalar field, but as
$K$ (and therefore $v_g$) becomes appreciable, the universe is
separated into different regions. In regions where the biscalar
field $\phi$ emerges close to the bottom of the potential, no
additional inflation occurs and presumably no (pre-)heating
occurs, and these regions remain largely empty of matter (other
than the radiation that is initially present in our model). In
regions where the biscalar field emerges farther up the potential,
early inflation may continue for a while before a period of
slow-roll-type inflation takes over, following which the universe
is (pre-)heated.

In the second period of inflation, the attractor solutions are not
necessarily slow-roll (although for models which generate
sufficient inflation, they effectively are during the time when
modes are generated and leave the horizon), rather they are
constant-roll, i.e., $\dot{\phi}\approx\mathrm{constant}$.  For
the biscalar field this has additional meaning: constant
$\dot{\phi}$ implies that $K\approx K_c$ is constant, and
therefore from~\eqref{eq:grav metric} and~\eqref{eq:g bar} the
speed at which gravitational waves travel is $v_g < c$. Therefore,
while the scalar mode spectrum alone is unlikely to be
distinguishable from some particular slow-roll model of inflation,
{\it the tensor modes have a different propagation velocity and new
physics is expected}.

\subsection{Solutions with Fixed Light Cones}
\label{sect:constant K}

We are looking for solutions that have a constant $K=K_c$ which
implies that $v_g < c$ is constant. From~\eqref{eq:K dot} this
implies that
\begin{equation}\label{eq:dot VB}
 3(1-K_c)H+K_c\dot{V}_B=0.
\end{equation}
Using~\eqref{eq:K cosm} to write this in terms of $K$ only
(assuming that $\dot{\phi}>0$) and~\eqref{eq:Fried eqn} to replace
$H$, this can be written as (assuming that $\Lambda=0$):
\begin{subequations}\label{eq:potential conditions}
\begin{equation}\label{eq:first condition}
\frac{1-K_c}{2K_c}+V_B=\frac{2K_c}{3(1-K_c)}(V_B^\prime)^2,
\end{equation}
which must hold at the beginning of the constant $K=K_c$ solution
which we define to begin at $\phi=\phi_c$ and $R=R_c$. Taking the
time derivative of (\ref{eq:first condition}), and defining the
dimensionless mass $m_c$ of the biscalar field near these
solutions, we find
\begin{equation}\label{eq:mc}
V_B^{\prime\prime}=\frac{3(1-K_c)}{4K_c}=m_c^2.
\end{equation}
\end{subequations}
Further time derivatives will require the vanishing of higher
order derivatives of the biscalar potential. The two
conditions~\eqref{eq:potential conditions} imply a quadratic
potential of the form:
\begin{equation}\label{eq:exact potential}
V_B=m_c^2\bigl[\tfrac{1}{2}(\phi-\phi_c)^2
-2\sqrt{2\tilde{N}_c}(\phi-\phi_c)+4N_c\bigr],\quad
\tilde{N}_c=N_c+\tfrac{1}{6},
\end{equation}
where $N_c$ is a parameter that will turn out to determine the number of e-folds of
inflation that  will take place if the biscalar field begins at $\phi_c$ in this
potential, and $\tilde{N}_c$ is introduced for convenience.

The condition~\eqref{eq:dot VB} determines the potential as a
function of the scale factor
\begin{equation}\label{eq:Kc potential}
V_B =4m_c^2\Bigl[N_c -\ln\Bigl(\frac{R}{R_c}\Bigr)\Bigr],
\end{equation}
where we have used the definition~\eqref{eq:mc} as well as
$V_B(\phi_c)=4m_c^2N_c$, which is consistent with the result
from~\eqref{eq:exact potential}. Using this, the Friedmann equation
becomes
\begin{equation}\label{eq:Kc Friedmann}
 H^2
 =6H_B^2(1-K_c)\Bigl[\tilde{N}_c-\ln\Bigl(\frac{R}{R_c}\Bigr)\Bigr],
\end{equation}
and the Hubble function at $t_c$ is given by
$H_c^2=6H_B^2(1-K_c)\tilde{N}_c$. Integrating this, we find the
solutions
\begin{equation}\label{eq:Rct}
R=R_c\exp\bigl[-\tfrac{3}{2}H_B^2(1-K_c)(t-t_c-2\tilde{N}_c/H_c)^2
+\tilde{N}_c\bigr],
\end{equation}
where an arbitrary sign has been chosen so that $R$ is increasing at
$R_c$.

This solution is Gaussian in time, and so it will reach a maximum
in $R$, following which it begins to contract. This cannot be
correct though, since at some point this means that the potential
has to become negative (recall that $H>0$ for $V_B>0$ and for $R$
to decrease then $H=0$ somewhere). If we define $R_f$ to be the
point at which $V_B=0$, then from the second form of~\eqref{eq:Kc
potential} we see that
\begin{equation}
R_f =R_ce^{N_c},
\end{equation}
which is the motivation for introducing the constant $N_c$ into
the potential. Although we shall not do so here, it is a
straightforward matter to show~\cite{Liddle+Lyth:2000} that the
solutions~\eqref{eq:Rct} are attractors, with nearby solutions
converging by a factor $\exp(-3N_c)$ over the period of inflation.

Evaluating the Hubble function at the same time gives
\begin{equation}\label{eq:Hf}
H_f^2=(1-K_c)H_B^2,
\end{equation}
using which, we can obtain
\begin{equation}
\frac{R^2_fH^2_f}{R^2_cH^2_c} =\frac{1}{6\tilde{N}_c}e^{2N_c}.
\end{equation}
Noting that the condition for sufficient inflation to solve the
horizon and flatness problems is (assuming instantaneous
(pre-)heating and standard physics, following
inflation~\cite{Coles+Lucchin:1995}):
\begin{equation}
\frac{R_c^2H_c^2}{R_0^2H_0^2}=
 \frac{1}{6\tilde{N}_c}e^{2N_c}
 \,10^{-60}z_{\mathit{dec}}\Bigl(\frac{T_P}{T_f}\Bigr)^2
 \gg 1,
\end{equation}
where we have assumed that following $R_f$ the universe matches on
to the standard big bang.
In this, $z_{\mathit{dec}}=4.3\times10^4\,\Omega\,h^2$ is the redshift at decoupling,
$0.4\le h =H_0/100\le 1$, $T_P=\sqrt{\hbar c^5/G}/k_{\mathrm{Boltz}}$ is the Planck
temperature, ($k_{\mathrm{Boltz}}$ is the Boltzmann constant), and $T_f$ is the
temperature at the end of inflation, where typically $T_f/T_P\in(10^{-5},1)$. This
gives the usual result that $N_c\gtrsim 60$ yields an adequate inflation factor to
solve the horizon and flatness
problems~\cite{Clayton+Moffat:2001,Coles+Lucchin:1995}. However, we note that since
this period is preceded by an earlier epoch of inflation, this is not the critical
issue. What is more important is that there be enough inflation for modes to be
created in an adiabatic vacuum state, which is the case if this is satisfied.

If we assume that (pre-)heating happens instantaneously, then
using~\eqref{eq:Hf} the energy density in the biscalar field $\phi$ is
transferred completely into the radiation field and we have
\begin{equation}
 \tfrac{1}{6}\kappa c^4 K_c^{\frac{3}{2}}\rho_f
 \approx (1-K_c)H_B^2.
\end{equation}
Relating this to the temperature at the end of inflation, we obtain
\begin{equation}\label{eq:temperature}
 \Bigl(\frac{T_P}{T_f}\Bigr)^2
 =\sqrt{\frac{\rho_P}{\rho_f}}
 \approx \frac{4\sqrt{2\pi}}{\xi_B}
 \frac{K_c^{\frac{3}{4}}}{\sqrt{1-K_c}},
\end{equation}
which determines the end of inflation given $K_c$ and $B$. We
observe that a larger value of $K_c$ tends to push the end of
inflation to lower temperatures, whereas if $K_c$ is small enough
then (pre-)heating will ``boost'' the radiation density suddenly,
and, depending on the parameters, the physics of (pre-)heating
could be drastically altered due to the presence of $\rho$ in
$\bar{\metric}_{\mu\nu}$ in \eqref{eq:g bar}.

Using $R_f$ and the solution for $R$, we can determine that during
this period of inflation the biscalar field rolls a distance
\begin{equation}
\Delta \phi
 =2\sqrt{2\tilde{N}_c}\Bigl(1-\frac{1}{\sqrt{6\tilde{N}_c}}\Bigr),
\end{equation}
and during this time it is rolling linearly in time
\begin{equation}
\phi =\phi_c+2H_B\sqrt{3(1-K_c)}(t-t_c).
\end{equation}
The scale factor can be written explicitly in terms of the biscalar field
\begin{equation}\label{eq:R phi}
R=R_c\exp\Bigl[-\frac{1}{8}(\phi-\phi_c)^2
+\sqrt{\frac{\tilde{N}_c}{2}}(\phi-\phi_c)\Bigr].
\end{equation}
Evaluating the potential explicitly in terms of the biscalar
field results in the potential~\eqref{eq:exact potential}. Note
though that there is no dependence on $m_c$ in~\eqref{eq:R phi},
so \textit{all} solutions with a given number $N_c$ of e-folds of
inflation will traverse the same path in the $R$-$\phi$ plane, but
will do so at potentially different rates, determined by $m_c$.

\section{Inflation Parameters}
\label{sect:parameters}

Using the solution obtained in the previous section, we can now
determine the inflation parameters. In the usual scenario, large
scale structure is linked to primordial quantum fluctuation of the
inflaton field. Following this, we consider quantum fluctuations
of the biscalar field $\phi$, and determine the resulting
spectrum. The analysis of linearized perturbations of the field
equations~\eqref{eq:Einsteins eqns} and~\eqref{eq:scalar FEQ} is
straightforward but rather involved, and we will merely quote the
required results, relegating most of the details to a future
publication.

The mechanism is centred on the
standard form of a quantum scalar field~\cite{Birrell+Davies:1989}:
\begin{subequations}\label{eq:quantum field}
\begin{equation}
\delta\hat{\phi}(t,\vec{x})=\int\frac{d^3k}{(2\pi)^{\frac{3}{2}}}\,
\Bigl[ a_{\vec{k}}\delta \tilde{\phi}e^{i\vec{k}\cdot\vec{x}}
+a^\dagger_{\vec{k}}\delta
\tilde{\phi}^*e^{-i\vec{k}\cdot\vec{x}}\Bigr],
\end{equation}
where the ladder operators satisfy:
$[a_{\vec{k}},a_{\vec{k}^\prime}]=0=
[a^\dagger_{\vec{k}},a^\dagger_{\vec{k}^\prime}]$, and
$[a_{\vec{k}},a^\dagger_{\vec{k}^\prime}]=\delta^3(\vec{k}-\vec{k}^\prime)$.
We require that these quantum field operators satisfy the equal
time commutation relations (note that here the scalar field is
dimensionless):
\begin{gather}
\label{eq:commutators}
[\delta\hat{\phi}(t,\vec{x}),\delta\hat{\phi}(t,\vec{x}^\prime)]=0=
[\delta\dot{\hat{\phi}}(t,\vec{x}),\delta\dot{\hat{\phi}}(t,\vec{x}^\prime)],\\
[\delta\hat{\phi}(t,\vec{x}),\delta\dot{\hat{\phi}}(t,\vec{x}^\prime)]
= \frac{i\hbar c^2\kappa\sqrt{K}}{R^3}\delta^3(\vec{x}-\vec{x}^\prime),
\end{gather}
\end{subequations}
where the additional factor of $K$ results from
the form of the metric ${\metric}_{\mu\nu}$.

The mode functions of a free
quantum field in an approximately flat spacetime (expanded in a
small time about some $R=R_i$) that satisfies our conditions would have the
form:
\begin{subequations}\label{eq:flat form}
\begin{equation}
\delta\tilde{\phi} = \sqrt{\frac{\hbar c^2 \kappa
\sqrt{K_c}}{2\omega_k R^3_i}} \bigl(C_+e^{-i\pi/4}e^{i\omega_k t}
+C_-e^{i\pi/4}e^{-i\omega_k t} \bigr),
%
\end{equation}
where
\begin{equation}\label{eq:vacuum}
|C_-|^2-|C_+|^2=1, \quad\rightarrow\quad
 C_-=1,\quad C_+=0,
\end{equation}
\end{subequations}
with the latter being the conventional adiabatic vacuum.

The second ingredient is the matched WKB approximation of the solution of an equation
of the form~\cite{Holmes:1995}:
\begin{subequations}
\label{eq:matched WKB}
\begin{equation}\label{eq:form}
 \partial_t^2\Sigma
 +U\Sigma=0,
\end{equation}
where there is a simple turning point at $t=t_k$ defined by $U(t_k)=0$, and $U>0$ for $t<t_k$, and $U<0$
for $t>t_k$.
In this case, the matched WKB approximation is given by
\begin{align}
 \tilde{\Sigma}_i&=|U|^{-\frac{1}{4}}
 \Bigl(\tilde{\Sigma}_i^+ e^{i\theta_i}
 +\tilde{\Sigma}_i^- e^{-i\theta_i} \Bigr),\quad
 t<t_k,\\
 \tilde{\Sigma}_o&=|U|^{-\frac{1}{4}}
 \Bigl(\tilde{\Sigma}_o^+e^{\theta_o}
 +\tilde{\Sigma}_o^-e^{-\theta_o}\Bigr),\quad
 t_k<t,
\end{align}
with
\begin{equation}
\theta_i=-\int_{t}^{t_k}dt\,\sqrt{U},\quad
\theta_o=\int^{t}_{t_k}dt\,\sqrt{-U}.
\end{equation}
The coefficients are related by
\begin{equation}\label{eq:matching}
\tilde{\Sigma}_0^+=e^{i\pi/4}\tilde{\Sigma}^+_i-e^{-i\pi/4}\tilde{\Sigma}^-_i,\quad
\tilde{\Sigma}_0^-=\tfrac{1}{2}e^{i\pi/4}\tilde{\Sigma}^-_i
-\tfrac{1}{2}e^{-i\pi/4}\tilde{\Sigma}^+_i.
\end{equation}
\end{subequations}

\subsection{Scalar Mode Perturbations}
\label{sect:scalar}

Considering scalar mode perturbations of the
system~\eqref{eq:Einsteins eqns}, it can be shown that
perturbations of the biscalar field can be determined from
\begin{equation}
\delta\phi=\frac{4}{\dot{\phi}\sqrt{R}}\bigl(\partial_t\Sigma+\tfrac{1}{2}H\Sigma\bigr),
\end{equation}
with $\Sigma$ satisfying an equation of the form~\eqref{eq:form} with
\begin{equation}
U=\frac{K_cc^2k^2}{R^2}+\frac{3}{2}\dot{H}-\frac{1}{4}H^2.
\end{equation}

Before the mode has left the horizon, the WKB solution~\eqref{eq:matched WKB} can be
matched to a flat spacetime solution of the form~\eqref{eq:flat form}, to give
\begin{equation}
\tilde{\Sigma}_i^\pm = \frac{\dot{\phi}}{4}
\sqrt{\frac{\hbar c^2 \kappa}{2c^2k^2\sqrt{K_c}}}
C_\pm e^{\mp i\pi/4}.
\end{equation}
Once the mode has moved outside the horizon, only the increasing mode of $\Sigma$
will be important, and using the above result with~\eqref{eq:matching}, we therefore
have
\begin{equation}
\tilde{\Sigma}_o\approx
\frac{\dot{\phi}}{4}\sqrt{\frac{\hbar c^2\kappa}{\sqrt{K_c}c^2k^2}}
\sqrt{\frac{R}{HR_k}}(C_+-C_-).
\end{equation}
In deriving this form, we have used the fact that very soon after the turning point
$U\approx(H/4)^2$, and so we obtain
\begin{equation}\label{eq:outside theta}
\theta_o\approx \frac{1}{2}\int_{t_k}^tdt\,H
=\frac{1}{2}\ln(R/R_k).
\end{equation}
Here, $R_k$ is defined by the turning point condition $U=0$ (using $H^2\gg \dot{H}$):
\begin{equation}\label{eq:turning point}
\frac{\sqrt{K_c}c^2k^2}{R_k^2}=\frac{1}{4}H_k^2.
\end{equation}
Near the turning point $\partial_t\Sigma \ll H\Sigma$, we find that
\begin{equation}
\delta\phi\approx \frac{4}{\dot{\phi}}H\Psi
=\sqrt{\frac{\hbar c^2\kappa}{\sqrt{K_c}c^2k^2}}
\sqrt{\frac{1}{HR_k}}(C_+-C_-).
\end{equation}

This result contains a few important features.
Since it ``runs'' slightly due to the presence of $H$, we merely evaluate it at the
end of inflation: $H\approx H_f$ defined in \eqref{eq:Hf}. This is not strictly
correct because near $R_f$, $\dot{H}$ is no longer negligible, but it will have the
correct order of magnitude. Also, because $H$ is not constant during inflation, the
solution of~\eqref{eq:turning point} does not just lead to $R_k\propto k$ and a
trivially scale invariant spectrum. Instead, we find that ($W(-1,x)$ is the branch of
the Lambert $W$-function appropriate for $-1\ll x<0$):
\begin{equation}
\Bigl(\frac{R_k}{R_c}\Bigr)^2
=e^{2\tilde{N}_c}\frac{-\lambda^2}{W(-1,-\lambda^2)}
\approx e^{2\tilde{N}_c}\frac{-\lambda^2}{\ln(\lambda^2)}.
\end{equation}
We have used the fact that (using $(4/3)\exp(-1/3)\approx 1$,
$4\sqrt{2\pi}\approx 10$ and~\eqref{eq:temperature}):
\begin{equation}
\lambda^2=\frac{4K_c}{3(1-K_c)}\frac{c^2k^2}{H_B^2R_c^2}e^{-2\tilde{N}_c}
\approx
 \frac{K_c^{\frac{7}{4}}}{\xi_B\sqrt{1-K_c}}
 \frac{c^2k^2}{R_0^2H_0^2}
\,10^{-59}z_{\mathrm{dec}},
\end{equation}
is small to keep only the asymptotic form of the solution.

We now have
\begin{equation}
R_k\approx
\frac{ck}{H_B}\sqrt{\frac{4K_c}{3(1-K_c)}}\frac{1}{\sqrt{-\ln(\lambda^2)}},
\end{equation}
which gives the scalar spectrum (evaluated at the end of inflation):
\begin{equation}\label{eq:scalar spectrum}
\mathcal{P}_{\delta\phi}
=\frac{k^3}{2\pi^2}|\delta\phi|^2
\approx
\frac{1}{\sqrt{3}\pi}
\xi_B^2
\frac{1-K_c}{K_c^{\frac{3}{4}}}[-\ln(\lambda^2)]
|C_+-C_-|^2.
\end{equation}

Following inflation and (pre-)heating, we expect that we can match this result to
standard post-inflation physics, and therefore using the fact that at the end of
inflation we have
\begin{equation}
\frac{H^2_f}{\dot{\phi}^2}\approx \frac{1}{12},
\end{equation}
the spectrum of curvature perturbations is determined by~\cite{Liddle+Lyth:2000}:
\begin{equation}
\mathcal{P}_R
=\frac{H^2}{\dot{\phi}^2}\mathcal{P}_{\delta\phi}
\approx \xi_B^2
\frac{1}{12\sqrt{3}\pi}
\frac{1-K_c}{K_c^{\frac{3}{4}}}[-\ln(\lambda^2)]
|C_+-C_-|^2.
\end{equation}
From this result, we can find the spectral
index~\cite{Liddle+Lyth:2000}:
\begin{equation}
 n_s=1+\frac{d\ln\mathcal{P}_{\mathcal{R}}}{d\ln k}
 =1+\frac{1}{\ln(\lambda)},\quad
 \alpha_s=\frac{d\,n_s}{d\ln k}=-\frac{1}{\ln(\lambda)^2}.
\end{equation}
Furthermore, we have
\begin{equation}\label{eq:delta H}
\delta_H
=\frac{2}{5}\sqrt{\mathcal{P}_R}
\approx\xi_B
\frac{2}{5\sqrt{12\sqrt{3}\pi}}
\frac{\sqrt{1-K_c}}{K_c^{\frac{3}{8}}}\sqrt{-\ln(\lambda^2)}
|C_+-C_-|.
\end{equation}

\subsection{Tensor Modes}
\label{sect:tensor}

The analysis of the tensor modes proceeds in the same way, although this time with
the tensor metric fluctuation: $\delta\metric_{ij}=h_{ij}=\delta\hat{\metric}_{ij}$,
and the transverse and traceless quantities: $\gamma^{ij}h_{ij}=0$ and
$\nabla^jh_{ij}=0$. The modes are represented in terms of the usual $I=\times,+$
polarization tensors $e_{ij}^I$~\cite{MTW:1973}:
\begin{equation}
h_{ij}=\frac{K^{\tfrac{1}{4}}}{R^{\frac{3}{2}}}e^I_{ij}\Sigma_I,
\end{equation}
where each of the $\Sigma_I$ satisfy an equation of the form~\eqref{eq:form} with
\begin{equation}
U=\frac{K_cc^2k^2}{R^2}-\frac{3}{2}\dot{H}-\frac{9}{4}H^2.
\end{equation}

The same procedure as described above leads to
\begin{equation}
\tilde{\Sigma}^{\pm}_i=\sqrt{\tfrac{1}{2}\hbar c^2\kappa}C_{\pm}e^{\mp
i\pi/4},
\end{equation}
where the $C_\pm$ again satisfy~\eqref{eq:vacuum}.
Outside the horizon we now have
\begin{equation}
 U\approx \tfrac{9}{4}H^2,\quad
 \theta_o \approx\tfrac{3}{2}\ln(R/R_k),
\end{equation}
and so well outside the horizon the modes have the form:
\begin{equation}
\tilde{\Sigma}_o \approx
\sqrt{\frac{\hbar c^2\kappa R^3}{3HR_k^3}}(C_+-C_-).
\end{equation}
Again these modes are slowly increasing, so we evaluate them at the end of
inflation~\eqref{eq:Hf}. In this case, the turning point is determined by
\begin{equation}\label{eq:tensor turning point}
\frac{\sqrt{K_c}c^2k^2}{R_k^2}=\frac{9}{4}H_k^2,
\end{equation}
which has the same solution as~\eqref{eq:turning point} with
$\lambda$ replaced by $3\lambda$.

Putting these results together, we find for each mode that
\begin{equation}\label{eq:P h}
\mathcal{P}_h
 =\frac{k^3}{2\pi^2}|h_{ij}|^2
 \approx  \xi_B^2
 \frac{3\sqrt{3}}{4\pi}
 \frac{1-K_c}{K_c}
 [-\ln(9\lambda^2)]^{\frac{3}{2}}
 |C_+-C_-|^2,
\end{equation}
using which we can determine the tensor spectral index:
\begin{equation}
 n_t
 =\frac{d\ln(\mathcal{P}_h)}{d\ln k}
 =\frac{3}{2\ln(3\lambda)},\quad
 \alpha_t=\frac{d\,n_t}{d\ln k} =-\frac{3}{2\ln(3\lambda)^2}.
\end{equation}
Finally, the ratio of the tensor to scalar mode contributions can
be determined using~\eqref{eq:delta H} and~\eqref{eq:P h}. To use
the latter result, we have ignored the running of $\lambda$ with
$k$ and we have taken $c_l\approx 1$ (see e.g., the discussion in
Section~7.7.2 of~\cite{Liddle+Lyth:2000}):
\begin{equation}
r\approx
\xi_B \sqrt{\frac{25\sqrt{6}}{8\pi}}
\Bigl(1+\frac{48\pi^2}{385}\Bigr)
\sqrt{1-K_c}K_c^{-\frac{5}{8}}
\frac{[-\ln(9\lambda^2)]^{\frac{3}{2}}}{[-\ln(\lambda^2)]^{\frac{1}{2}}}.
\end{equation}

\subsection{Observational Constraints}
\label{sect:constraints}

Assuming an adiabatic vacuum, $C_+=0$ and $C_-=1$,
$z_{\mathit{dec}}=4.0\times 10^4$, and evaluating everything at
the pivot point $ck_{\mathrm{pivot}}=7\,R_0H_0$,
from~\eqref{eq:delta H} we obtain~\cite{Liddle+Lyth:2000}:
\begin{subequations}
\label{eq:fit deltaH}
\begin{equation}
\delta_H
\approx \xi_B \frac{\sqrt{1-K_c}}{20K_c^{\frac{3}{8}}}\sqrt{-\ln(\lambda^2)}
\approx 1.91\times 10^{-5},
\end{equation}
for
\begin{equation}
\lambda^2
\approx 2\times 10^{-53}\,\frac{K_c^{\frac{7}{4}}}{\xi_B\sqrt{1-K_c}}.
\end{equation}
\end{subequations}
Satisfying~\eqref{eq:fit deltaH} allows one to determine $\xi_B$
in terms of $K_c$. Once this is done, we find that we can restrict
$K_c$ to lie in the range
\begin{equation}\label{eq:Kc limits}
K_c\in[2.0\times 10^{-15},1.0-1.3\times 10^{-13}],
\end{equation}
by requiring that $T_f\le T_P$ (giving the lower limit) and
$\xi_B\le 32\pi$ (giving the upper limit). That is, we require
that the temperature at the end of inflation is below the Planck
temperature and the fundamental length $\sqrt{B}$ appearing
in~\eqref{eq:bimetric} is not less than the Planck length. As we
have discussed
elsewhere~\cite{Clayton+Moffat:1999a,Clayton+Moffat:2001}, it is
far from clear how to define a fundamental scale that plays the
role of the Planck length in this type of model.  Nevertheless,
these seem like reasonable conditions.

Over this range, several of the parameters are rather insensitive:
\begin{subequations}
\begin{gather}\label{eq:constants}
n_s\in[0.982, 0.987],\quad
n_t\in[-0.027,-0.019],\\
\alpha_s\in[-3.2,-1.6]\times 10^{-4},\quad
\alpha_t\in[-5.0,-2.5]\times 10^{-4}.
\end{gather}
\end{subequations}
In all cases, the lower limit corresponds to the upper limit of
$K_c$. Note that the very small tilt rules out cold dark matter
models within this scenario~\cite{White+etal:1995}, and the small
values of $\alpha_s$ and $\alpha_t$ imply that any running of
the spectral indices should be unobservable in data from the MAP
and Plank satellite missions~\cite{Copeland+Grivell+Liddle:1997}.

The remaining parameters display similar trends, all of them being
roughly constant for $K_c\gtrsim 10^{-7}$ with the values
\begin{subequations}
\begin{equation}\label{eq:lower}
r\approx 0.014,\quad
T_f\approx 0.0019\,T_P,\quad
m_c\xi_B\approx 0.0018.
\end{equation}
They all possess exponential behavior for smaller $K_c$, and at
the lower limit of~\eqref{eq:Kc limits} have the values:
\begin{equation}
r\approx 77,\quad
T_f\approx T_P,\quad
m_c\xi_B\approx 3.1\times 10^{-5}.
\end{equation}
\end{subequations}
The combination $m_c\xi_B$ determines the length scale appearing
in the biscalar potential, from which we find that the biscalar
field mass is $m_c=2.2\times 10^{16}\,\mathrm{GeV}$, decreasing to
$m_c=3.8\times 10^{14}\,\mathrm{GeV}$ at the lower limit
of~\eqref{eq:Kc limits}. The observation of CMB polarization from
the MAP and Planck satellite missions~\cite{Kinney:1998} should
constrain $r$, effectively fixing the remaining parameters of the
model. Note the lack of a simple `consistency relation' between
$r$ and $n_s$ that is typical of inflaton models. For a quadratic
inflaton potential one would have $r=7(1-n_s)/2$, and so even in
the small $K_c$ limit the tensor modes are larger by a factor
$\approx 2$.

The most complicated behavior is displayed by $\xi_B$, which near
$K_c\approx 1$ is increasing exponentially to reach $\xi_B=32\pi $
at the upper limit in~\eqref{eq:Kc limits}. In the intermediate
regions, it is relatively constant at $\xi_B\approx
\exp(-10)\approx 4.5\times 10^{-5}$, and then approaches its lower
limit $\xi_B\approx 10^{-10}$ at the lower limit of $K_c$. If we
choose $B=1/\ell_P^2$ as suggested in~\cite{Clayton+Moffat:2001},
then we are in the regime where $\xi_B$ is approximately constant,
and we find the parameter set:
\begin{subequations}
\begin{gather}
1-K_c\approx 1.3\times 10^{-9},\quad m_c\xi_B\approx
0.0018 ,\quad \xi_B\approx 1,\quad T_f/T_P\approx 0.0019 ,\\ n_s\approx 0.982,\quad
n_t\approx -0.027,\quad \alpha_s\approx-3.2\times 10^{-4},\quad
\alpha_t\approx-5.0\times 10^{-4},\quad r\approx0.014.
\end{gather}
\end{subequations}
These predictions are consistent with
observational results~\cite{Lyth+Riotto:1999} and in particular
with the recent MAXIMA
results~\cite{Balbi+etal:2000,Stompor+etal:2001}.

\section{Conclusions}
\label{sect:conclusion}

We have investigated a model of the early universe based on a
bimetric gravitational theory with a biscalar field $\phi$, which
has an initial period of inflation when the speed of gravitational
waves $v_g\rightarrow 0$. This period is insensitive to the
potential $V(\phi)$ and to the initial values of $\phi$. Two
possible models for the generation of vacuum fluctuations present
themselves: (i) the biscalar field $\phi$ both initiates inflation
and its quantum vacuum fluctuations serve to generate the CMB
spectrum after decoupling, (ii) the biscalar field $\phi$ produces
an initial period of inflation as $v_g\rightarrow 0$ but the
quantum vacuum fluctuations of a matter scalar field $\psi$, on
the initial inflating background, generates the seeds of the CMB
spectrum after the fluctuations cross the horizon.

We have concentrated on calculating the predictions of model (i),
in which a second period of inflation begins when $v_g$ increases
towards the measured speed of light $c$ and the potential
$V(\phi)$ begins to play a significant role. This second period of
inflation is characterized by a ``constant-roll'' of $\phi$ towards
the bottom of the potential. A calculation yields predictions for
the parameters describing the CMB spectrum, which agree with
current observations.  In a subsequent paper, we will obtain
predictions for model (ii) and we anticipate that in this scenario
a fine tuning of the $\psi$ scalar field mass and coupling
constant will not occur, in contrast to the standard slow-roll
inflaton models of inflation.

A significant aspect of both models (i) and (ii) is that the initial period
of inflation, mediated by $v_g\rightarrow 0$, suggests a lack of sensitivity to any
potential transplanckian physics influence.  Such an
influence could lead to a lack of predictability of the early universe cosmologies,
as appears to be the case for standard slow-roll inflationary
models~\cite{Brandenberger+Martin:2002}. It is also important to note that the predictions of
BGT for the scalar-tensor fluctuation modes do not necessarily coincide with the
standard inflaton inflationary models. Perhaps, future observations of the CMB
spectrum, using more accurate data, will be able to distinguish between the BGT
predictions and other inflationary scenarios.

We are now in a position to claim that bimetric gravitational
theories can produce predictions in agreement with the CMB
spectrum observations, and therefore provide an alternative
scenario to the standard slow-roll inflationary models.

\section*{Acknowledgments}

This work was supported by the Natural Science and Engineering Research
Council of Canada.


\end{document}